\documentclass[letter,scriptaddress,twocolumn, prl,showkeys,showpacs,superscriptaddress]{revtex4-1}
\usepackage[toc]{appendix}
	\usepackage{amsmath}
	\usepackage{mathtools}
	\usepackage{makeidx}
	\usepackage{amsfonts}
	\usepackage{graphicx, xcolor}  
	\usepackage{dcolumn}   
	\usepackage{bm}        
	\usepackage{amssymb}   
	\usepackage[ansinew]{inputenc}
	\usepackage[usenames,dvipsnames]{pstricks}
	\usepackage{subfigure}
	\usepackage{epstopdf}
	\usepackage{pst-grad} 
	\usepackage{pst-plot} 
	\usepackage[colorlinks,hyperindex]{hyperref}
	\usepackage{tikz}
	\hypersetup
	{
		colorlinks,%
		citecolor=black,%
		linkcolor=black,%
		urlcolor=black,%
	}
	\usepackage{auto-pst-pdf}
	
\newcommand{\be}{\begin{equation}}
\newcommand{\ee}{\end{equation}}
\newcommand{\ba}{\begin{array}}
\newcommand{\ea}{\end{array}}
\newcommand{\bml}{\begin{multline}}
\newcommand{\eml}{\end{multline}}

\newcommand{\mh}{\mathcal{H}}

\newcommand{\mf}{\mathcal{F}}

\newcommand{\rd}{{\rm d}}

\newcommand{\dH}{\Delta\mathcal{H}}
\newcommand{\mjs}{\langle Jss\rangle_{m'}}
\newcommand{\ml}{\langle\lambda\rangle_{m'}}

\newcommand{\lp}{\left(}
\newcommand{\rp}{\right)}
\newcommand{\lm}{\lambda_{m'}}

\begin{document}

\title{Dynamics and Correlations among Soft Excitations in Marginally Stable Glasses}
\author{Le Yan\thanks{ly452@nyu.edu}}
\affiliation{Center for Soft Matter Research, Department of Physics, New York University, \\4 Washington Place, New York, 10003, NY}
\author{Marco Baity-Jesi\thanks{marcobaityjesi@gmail.com}}
\affiliation{Departamento de F\'isica Te\'orica I, Universidad Complutense, 28040 Madrid, Spain}
\affiliation{Dipartimento di Fisica, La Sapienza Universit\`a di Roma, 00185 Roma, Italy}
\affiliation{Instituto de Biocomputaci\'on y F\'isica de Sistemas Complejos (BIFI), 50009
Zaragoza, Spain}
\author{Markus M\"uller\thanks{markusm@ictp.it}}
\affiliation{The Abdus Salam International Center for Theoretical Physics, \\Strada Costiera 11, 34151 Trieste, Italy}
\author{Matthieu Wyart\thanks{mw135@nyu.edu}}
\affiliation{Center for Soft Matter Research, Department of Physics, New York University, \\4 Washington Place, New York, 10003, NY}

\date{\today}

\begin{abstract}


Marginal stability is the notion that stability is achieved, but only barely so. This property constrains the ensemble of configurations explored at low temperature in a variety of systems, including spin, electron and structural glasses. A key feature of marginal states is a (saturated) pseudo-gap in the distribution of soft excitations. We study how such a pseudo-gap appears dynamically in the case of the Sherrington-Kirkpatrick (SK) spin glass. After revisiting and correcting the multi-spin-flip criterion for local stability, we show that stationarity along the hysteresis loop requires that soft spins are frustrated among each other, with a correlation that diverges as $C(\lambda)\sim 1/\lambda$, where $\lambda$ is the larger of two considered local fields. We explain how this arises spontaneously in a marginal system and develop an analogy between the spin dynamics in the SK model and random walks in two dimensions. We discuss the applicability of these findings to hard sphere packings. 
\end{abstract}

\maketitle

{\it Introduction:} In glassy materials with sufficiently long-range interactions, stability at low temperature imposes an upper bound on the density of soft excitations~\cite{Muller14}. 
In electron glasses~\cite{Efros75,Pollak13,
Muller04,Muller07,Palassini12,Andresen13a} stability toward hops of individual localized electrons requires that the density of states vanishes at the Fermi level, exhibiting a so-called Coulomb gap. Likewise, in mean-field spin glasses ~\cite{Thouless77,Pazmandi99,Pankov06,Eastham06,Horner07,Doussal10,Le-Doussal12,Andresen13
} stability towards flipping several ``soft" spins implies that the distribution of local fields vanishes at least linearly. In hard sphere packings the distribution of forces between particles in contact must  vanish analogously, preventing that collective motions of particles lead to denser packings~\cite{Wyart12,Lerner13a,DeGiuli14b
}. Often, these stability bounds appear to be saturated~\cite{Pazmandi99,Palassini12,Andresen13,Lerner13a,Lerner12,Charbonneau14c}.  Such {\em marginal stability} can  be proven for dynamical, out-of-equilibrium situations  under slow driving at zero temperature \cite{Muller14} if the effective interactions do not decay with distance. This situation occurs in the {Sherrington-Kirkpatrick (SK) model [defined below in Eq.~(\ref{h})]}, but also in finite-dimensional hard sphere glasses, where elasticity induces non-decaying interactions~\cite{Wyart05b}.  This result is also obtained for the ground state or for slow thermal quench by replica calculations in spin glass \cite{Mezard87,Pankov06} and hard sphere systems~\cite{Charbonneau14,Charbonneau14a} assuming infinite dimension.

The presence of pseudo-gaps strongly affects the physical properties of these glasses. The Coulomb gap alters transport properties in disordered insulators \cite{Efros75,Pollak13}, while its cousin in spin glasses suppresses the specific heat and susceptibility. It was recently proposed that the singular rheological properties of dense granular and suspension flows near jamming are controlled by the pseudo-gap exponents in these systems \cite{DeGiuli14d}. More generally, an argument of Ref.~\cite{Muller14} shows
that a pseudo-gap implies avalanche-type response to a slow external driving force, so-called crackling~\cite{Sethna01}. This is indeed observed in these systems \cite{Combe00,Palassini12,Pazmandi99}.  Despite the central role of pseudo-gaps, it has not been understood how they emerge dynamically, even though some important elements of the athermal dynamics of the SK spin glass have been pointed out in earlier works~\cite{Eastham06,Horner07}. 

In this Letter we identify a crucial ingredient that was neglected in previous dynamical approaches, and also in considerations of multi-spin stability: Soft spins are strongly frustrated among each other, a correlation that becomes nearly maximal  for spins in the weakest fields. 
These correlations require revisiting earlier multi-spin stability arguments that assumed opposite correlations. We then argue, assuming stationarity along the hysteresis loop, that the correlation $C(\lambda)$ between the softest spins and spins in local fields of magnitude $\lambda$ must follow $C(\lambda)\sim 1/\lambda^\gamma$, with $\gamma=1$. 
This is built into a Fokker-Planck description of the dynamics, which is used to predict the statistics of the number of times a given spin flips { in an avalanche. 

{\it Model:}  We consider the SK model with $N$ Ising spins ($s_i=\pm1$):
\be
\label{h}
\mh=-\frac{1}{2}\sum_{i\neq j}J_{ij}s_is_j-h\sum_{i=1}^Ns_i,
\ee
in an external field $h$. Every spin couples to all other spins $j$ via a symmetric matrix $J_{ij}$, whose elements are {\it i.i.d.} {Gaussian random variables} with mean $0$ and variance $1/N$. 
The total magnetization is $M\equiv\sum_is_i$. We define the local field $h_i$ and the local stability $\lambda_i$ of spin $i$ by
\be
\label{lambda}
h_i\equiv -\frac{\partial \mh}{\partial s_i} = \sum_{j\neq i}J_{ij}s_j+h,\quad\lambda_i=h_is_i.
\ee
A spin is called stable when it aligns with the local field, $\lambda>0$, and unstable otherwise. The energy to  flip the spin $s_i\to-s_i$ (and hence $\lambda_i\to-\lambda_i$) is:
\be
\label{e1}
\Delta\mh_1(i)\equiv\mh(-s_i)-\mh\\=2s_i(\sum_{j\neq i}J_{ij}s_j+h)=2\lambda_i. 
\ee


As in Ref.~\cite{Pazmandi99}, we consider the hysteresis loop at zero temperature obtained by quasi-statically increasing the field, as shown in Fig.~\ref{skmodel}(a).  When a spin turns unstable, we apply a greedy Glauber dynamics that relaxes the system in an avalanche-like process toward a new one-spin-flip stable state by sequentially flipping the most unstable spin.   
Those states empirically exhibit a pseudo-gap in the distribution of the $\lambda_i$ ~\cite{Pazmandi99,Eastham06,Andresen13},
\be
\label{pseudogap}
\rho(\lambda)=A\lambda^{\theta}+O(N^{-\theta/(1+\theta)}),
\ee
with $\theta=1$ for $\lambda\ll1$, as shown in Fig.~\ref{skmodel}(c), {but with a slope $A$ significantly larger than in equilibrium~\cite{Parisi03,Pankov06,Horner07}}. 
The avalanche size is power-law distributed~\cite{Pazmandi99}:
\be
\label{dn}
D(n)=n^{-\tau}d(n/N^\sigma)/\Xi(N),
\ee 
where $n$ is the number of flips in an avalanche. The scaling function $d(x)$ vanishes for $x\gg1$. $N^{\sigma}$ is the finite size cutoff, and $\Xi(N)$ is a size dependent normalization if $\tau\leq1$. 
Numerical studies of the dynamics of the SK model indicate that $\tau=\sigma=1$ and $\Xi=\ln N$~\cite{Pazmandi99,Andresen13}, as shown by the finite size collapse in Fig.~\ref{skmodel}(b). While one can argue  that $\theta=1$ along the hysteresis curve~\cite{Muller14}, the exponents $\tau$ and $\sigma$  have not been derived theoretically for the dynamics (unlike for ``equilibrium avalanches", for which $\tau=1$ has been obtained analytically~\cite{Doussal10,Le-Doussal12}). 

Below we present a theoretical analysis of the dynamics, making the following two assumptions:

{\it (i)} Contained Avalanches: Spins never become strongly unstable, that is, the lowest local stability encountered in an avalanche, $\lambda_0$, satisfies $\lambda_0\rightarrow 0$ as $N\rightarrow \infty$. This has been numerically verified in Fig.~\ref{propt}(a). 

{\it (ii)}  Stationarity: 
The number of times each spin flips along the hysteresis loop diverges with $N$ for any finite interval of applied field $[h, h+\Delta h]$ if $h=O(1)$. This assures that a stationary regime is reached very quickly.
(For $\tau=1$ this condition simply reads $\sigma +1/(1+\theta)>1$)~\footnote{The typical external field increment triggering an avalanche is $h_{\rm min}\sim\lambda_{\rm min}\sim N^{-1/(1+\theta)}$, so there are $N_{\rm av}\sim1/h_{\rm min}\sim N^{1/(1+\theta)}$ avalanches in a finite range of external field $dh$~\cite{Muller14}. Each avalanche contains on average $N_{\rm flip}\sim\int nD(n)\rd n\sim N^{(2-\tau)\sigma}$ flip events. The total number of flip events along the hysteresis curve is $N_{\rm av}N_{\rm flip}\sim N^{(2-\tau)\sigma+1/(1+\theta)}$, which we assume to be $\gg N$.}.

\begin{figure}[h!]
\includegraphics[width=1.0\columnwidth]{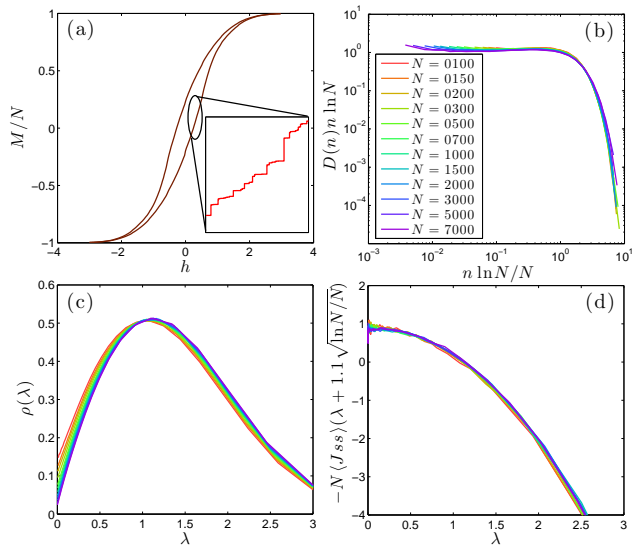}
\caption{\small{(a) Hysteresis loop: Magnetization $M$ under a periodic quasi-static driving of the external field $h$. Inset: magnified segment of the hysteresis loop of a finite size system. (b) Finite size scaling of avalanche size distribution $D(n)$ confirms $\tau=\sigma=1$ {up to logarithmic corrections}. 
(c) Distribution of local stabilities, $\rho(\lambda)$, in locally stable states along the hysteresis loops for different system sizes $N$. (d) Correlation $-\langle Jss\rangle$ between the least stable spin and spins at local stability $\lambda$ in locally stable states along the hysteresis loop. 
The data for different system sizes collapses.
}}\label{skmodel}
\end{figure}
 
{\it Multi-spin stability criterion:}
A static bound for the pseudogap exponent $\theta$ is obtained by considering two of the softest spins $i,j$ (with stabilities $\lambda_{\rm min}\sim 1/N^{1/(1+\theta)}$) ~\cite{Palmer79,Anderson79,Muller14}. Their simultaneous flip costs an energy $2(\lambda_i+\lambda_j - 2 s_is_j J_{ij})$. 
The last term scales as $1/\sqrt{N}$ and is negative if the two spins are {\em unfrustrated}. If this occurs with finite probability, a strong enough pseudogap, $\theta\geq 1$, is necessary to prevent the last term from overwhelming the stabilizing terms. However, extending this argument to  multi-spin stability
reveals its rather subtle nature.  Flipping a set $\mf$ of $m$ spins in an initial one-spin flip stable state costs an energy
\be
\label{de}
\Delta\mh(\mf)=2\sum_{i\in\mf}\lambda_i-2\sum_{i,j\in\mf}J_{ij}s_is_j.
\ee
The initial state is unstable to multi-flip excitations if $\Delta\mh<0$ for some $\mf$. Refs.~\cite{Palmer79, Anderson79}  considered just the set of the $m$ softest spins. Extremal statistics and the assumption of Eq.~(\ref{pseudogap}) implies the scaling of the maximal stabilities $\lambda(m)\sim (m/N)^{1/(1+\theta)}$, and thus $\sum_{i\leq m}\lambda_i\sim m \lambda(m)$.
The term $\sum_{i\leq m}J_{ij}s_is_j\sim m(m/N)^{1/2}$ was (erroneously) argued to be  positive {\em on average}, which yielded the bound $\theta\geq 1$ to guarantee $\Delta\mh_m>0$. 
However, numerically we find that on average $\sum_{i\leq m}J_{ij}s_is_j$ is negative for soft spins. More precisely, the correlation $C(\lambda)=-2\langle Jss\rangle$ between a spin of stability $\lambda$ and the softest spin in the system is positive for small $\lambda$, as shown in  Fig.~\ref{skmodel}(d). Postulating that:
\be
\label{correlation}
C(\lambda)\sim \lambda^{-\gamma} N^{-\delta},
\ee
it is straightforward to estimate that $\langle -\sum_{i\leq m}J_{ij}s_is_j \rangle\sim m^2 C(\lambda(m))\sim m^{2-\gamma/(1+\theta)} N^{\gamma/(1+\theta)-\delta}$. { A more complete characterization of correlations is given in the Supplementary Materials Section A and B.}

It follows that the average r.h.s. of Eq.~(\ref{de}) is always positive. We argue that the stability condition nevertheless leads to a non-trivial constraint, because the last term of Eq.~(\ref{de}) can have large fluctuations. Indeed, consider all sets $\mf$ of $m$ spins belonging to the $m'>m$ softest spins. For definiteness we choose $m'=2m$ and  use scaling arguments. A more detailed derivation for a general ratio $m'/m$ is given in the Supplementary Materials Section A. To estimate the probability that the optimal set leads to a negative $\Delta \mh$ in Eq.~(\ref{de}), we use a type of argument as often applied in the random energy model~\cite{Derrida81}. We assume that the  $\Delta\mh$ associated with different $\mf$ are independent, Gaussian distributed variables. The variance of the fluctuation $X\equiv \sum_{i,j\in\mf}J_{ij}s_is_j-\langle \sum_{i,j\in\mf}J_{ij}s_is_j\rangle$ is of order $m/\sqrt{N}$. Since there are $2^{2m}$ sets $\mf$, the number density having fluctuation $X$ follows ${\cal N}(X)\sim \exp[2m\ln(2)-X^2N/m^2]$. The most negative fluctuation $X_{\rm min}$ is determined by ${\cal N}(X_{\rm min})\sim 1$ leading to $X_{\rm min}\sim - m^{3/2}/\sqrt{N}$.  The associated energy change is thus, according to Eq.~(\ref{de}) and the subsequent estimates of each term:
\begin{multline}
\label{sta}
\Delta\mh(\mf_{\rm min})=m^{(2+\theta)/(1+\theta)}/N^{1/(1+\theta)}+ \\
m^{2-\gamma/(1+\theta)} N^{\gamma/(1+\theta)-\delta}- m^{3/2}/\sqrt{N}.
\end{multline}
Multi-spin stability requires that for large $N$ and fixed $m$ this expression be positive. This yields the conditions:
\be
\label{sca}
\theta\geq 1, \ \ \hbox{ or}\ \ \gamma/(1+\theta)-\delta\geq-1/2.
\ee
However, the correlation in Eq.~(\ref{correlation}) cannot exceed the typical coupling among spins, $C\lesssim1/\sqrt{N}$, which requires $\gamma/(1+\theta)-\delta\leq -1/2$. Thus, if $\theta<1$, stability imposes
the equality $\gamma/(1+\theta)-\delta=-1/2$, while the scaling with $m\gg 1$ additionally requires $2-\gamma/(1+\theta)\geq 3/2$; or in other words, $\gamma\leq(1+\theta)/2\leq1$ and $\delta\leq1$. 
In the relevant  states, all three exponents $\theta$, $\gamma$, and $\delta$ turn out to equal $1$ and thus satisfy these constraints as exact equalities.
We will now show how to understand this fact from a dynamical viewpoint.

\begin{figure}[h!]
 \centering
   \def\svgwidth{1.0\columnwidth}
   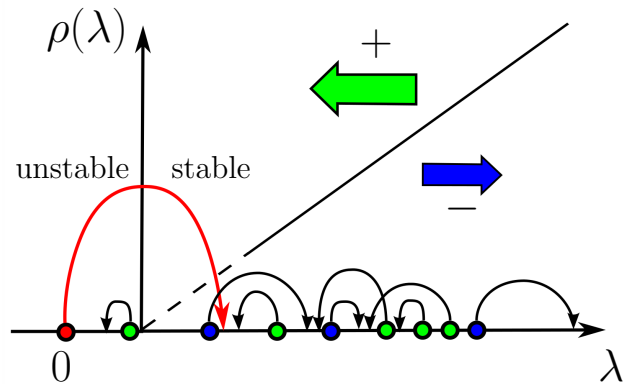
  \caption{\small{Illustration of a step in the dynamics, in the SK model and the random walker model. Circles on the $\lambda$-axis represent the spins or walkers. At each step, the most unstable spin or walker (in red) is reflected to the stable side, while all other spins or walkers (in green or blue) receive a kick and move. The dashed and solid line outlines the density profile  $\rho(\lambda)\sim\lambda$ for $\lambda>1/\sqrt{N}$. 
 The blue spins  were initially frustrated with the flipping spin 0. They are stabilized and are now unfrustrated with 0. In contrast, green spins become are frustrated with spin 0 and are softer now. Because of the opposite motion of spins according to their frustration with spin 0, a correlation builds up at small $\lambda$, leading to an overall frustration of ``soft" spins among each other.
 }}\label{dynmod}
\end{figure}

{\it Fokker-Planck equation:}   Consider an elementary spin flip event in the greedy relaxation dynamics, cf. Fig.~\ref{dynmod}. The local stability of the flipping spin (colored red and labelled $0$) changes from $\lambda_0$ to $-\lambda_0$ as the spin flips from $s_0$ to $-s_0$. Due to the coupling $J_{0j}$, the local stability of all other spins $j$, colored green or blue, changes accordingly,
\be
\label{dl}
\lambda_j\to\lambda_j'=\lambda_j-2J_{0j}s_0s_j.
\ee
These kicks have a random fluctuating part, as well as a mean value due to the correlation between the spins $0$ and $j$. Following \cite{Eastham06,Horner07} we model this dynamics via a Fokker-Planck equation for the distribution of local stabilities $\rho(\lambda,t)$: 
\begin{multline}
\label{fp}
\partial_t\rho(\lambda,t)=-\partial_{\lambda}\,\left[v(\lambda,t)-\partial_{\lambda}D(\lambda,t)\right]\rho(\lambda,t)\\-\delta(\lambda-\lambda_0)+\delta(\lambda+\lambda_0),
\end{multline}
where $t$ is the number of {\em flips  per spin} that have taken place.
$v(\lambda,t)\equiv-2N\langle J_{0i}s_0s_i\rangle_{\lambda_i = \lambda}\equiv NC(\lambda,t)$ is the average drift, and $D(\lambda,t)\equiv N \left(4\langle J_{0i}^2\rangle-[v(\lambda,t)/N]^2\right)/2$ is the diffusion constant. For $\lambda\gg 1/\sqrt{N}$,  $D=2$. 

For the dynamics to have a non-trivial thermodynamic limit the scaling $v\sim O(1)$ or $\langle J_{0i}s_0s_i\rangle\sim 1/N$ must hold, i.e., $\delta=1$ in Eq.~(\ref{correlation}).

Next we use our assumption {\em (i)}  that $\lambda_0\rightarrow 0$ in the thermodynamic limit. We may then replace the delta-functions in Eq.~(\ref{fp}) by a reflecting boundary condition at $\lambda=0$,
\be
\label{refb}
\left. \left[v(\lambda,t)-\partial_{\lambda}D(\lambda,t)\right]\rho(\lambda,t)\right|_{\lambda=0}=0.
\ee

Our stationarity assumption {\em (ii)} implies that finite intervals $dh$ along the hysteresis loop correspond to diverging times $\Delta t\rightarrow \infty$, that is, a diverging number of flips per spin. 
At those large times a dynamical steady state (ss) must be reached. In such a state the flux of spins must vanish everywhere:
\be
\label{steady}
v_{\rm ss}(\lambda)=D\partial_{\lambda}\rho_{\rm ss}(\lambda)/\rho_{\rm ss}(\lambda)\rightarrow 2\theta/\lambda
\ee
A similar result was obtained in Ref.~\cite{Horner07} following a quench.

{\it Emergence of correlations:}   Eq.~(\ref{steady}) implies that $\gamma=1$  in Eq.~(\ref{correlation}),  assuming Eq.~(\ref{pseudogap}) for $\rho_{\rm ss}$. Such singular correlations are unexplained \footnote[0]{ The approximation Eq.~(21) in Horner yields an incorrect scaling behavior for $C(\lambda)$, assuming a pseudogap.}. We now argue that they naturally build up in the dynamics through the spin-flip induced motion of stabilities of frustrated and unfrustrated spins, as illustrated in Fig.~(\ref{dynmod}).
To quantify this effect we define respectively $C_f(\lambda)$ and $C_f'(\lambda)$ as the correlation between the flipping spin 0 and the spins at $\lambda$ {\it before} and {\it after} a  flip event. 
The quantity $x_i\equiv-2J_{0i}s_0s_i$ is the kick in the stability of spin $i$ due to the flip of $s_0$, $\lambda'_i=\lambda_i+x_i$. 
The correlation $C_f'(\lambda)$ is the sum over the contributions from all spins which migrated to $\lambda$ due to the flip:
 \[
\begin{aligned}
C_f'(\lambda)&=\frac{1}{\rho'(\lambda)}\int\rho(\lambda-x)(-x)f_{\lambda-x}(x)\rd x,\\
\rho'(\lambda)&=\int\rho(\lambda-x)f_{\lambda-x}(x)\rd x.
\end{aligned}
\]
Here $f_{\lambda}(x)$ is a Gaussian distribution of the random variable $x$ at $\lambda$: $f_{\lambda}(x)=\exp\left[-\frac{(x-C_f(\lambda))^2}{4D/N}\right]/{\sqrt{4\pi D/N}}$. 
%
Expanding in the integrands $\rho(\lambda-x)$ and $C_f(\lambda-x)$ for small $x$ and keeping terms of order $1/N$, we obtain 
\begin{subequations}
\begin{align}
C_f'(\lambda)&=-C_f(\lambda)+2\frac{D}{N}\frac{\partial_{\lambda}\rho(\lambda)}{\rho(\lambda)},
\label{selfa}\\
\rho'(\lambda)&=\rho(\lambda)-\partial_{\lambda}\left[ C_f(\lambda)\rho(\lambda)-\frac{D}{N}\partial_{\lambda}\rho(\lambda)\right]. \label{selfb}
\end{align}
\end{subequations}
Thus,  even if correlations are initially absent ($C_f(\lambda)=0$),
they arise spontaneously, $C_f'(\lambda)=2D\partial_\lambda\rho(\lambda)/N\rho(\lambda)$. 

In  the steady state, $\rho'_{\rm ss}=\rho_{\rm ss}$. Eq.~(\ref{selfb}) then implies the vanishing of the spin flux, i.e., Eq.~(\ref{steady}) with $v=NC_f$.
Plugged into Eq.~(\ref{selfa}), we obtain that the correlations are steady, too, 
\be
\label{cc}
C'_f(\lambda)=C_f(\lambda)=\frac{v_{\rm ss}(\lambda)}{N}
=\frac{2\theta}{N\lambda}.
\ee 
These correlations are expected once the quasi-statically driven dynamics reaches a statistically steady regime, and thus should be present both
during avalanches and in the locally stable states reached at their end. 

Interestingly, Eq.~(\ref{cc})  implies that all the bounds of Eq.~(\ref{sca}) are saturated if the first one is, i.e., if $\theta=1$. {The latter value was previously derived from dynamical considerations} in Ref.~\cite{Muller14}. It is intriguing, that the present Fokker-Planck description of the dynamics  does not pin  $\theta$, as according to Eqs.~(\ref{steady},~\ref{cc}) any value of $\theta$ is acceptable for stationary states. However, additional considerations on the applicability of the Fokker-Planck description discard the cases $\theta>1$ and $\theta<1$, as discussed in the Supplementary Materials Section C. 


It is interesting to observe that Eqs.~(\ref{fp},~\ref{refb},~\ref{steady}) with $\theta=1$ are equivalent to the Fokker-Planck equation for the radial component of a  two-dimensional unbiased diffusion (the analogy is derived in Supplementary Materials Section D), whose statistics is well known \cite{redner01,Bray13}. We can use this analogy to predict $F(n)$, the number of times an {initially soft} spin flips in an avalanche of size $n$. Indeed, a discrete random walker starting at the origin will visit that point $\ln(t)$ times after $t$ steps in two dimensions, thus $F(n)\sim\ln(n)$. This prediction is confirmed in Fig.~\ref{propt}(b). Using the same analogy, we predict that the distribution of times between successive flips of a given spin goes as $P(\delta t) \sim 1/( \delta t [\ln(\delta t)]^2)$.

\begin{figure}[h!]
\includegraphics[width=1.0\columnwidth]{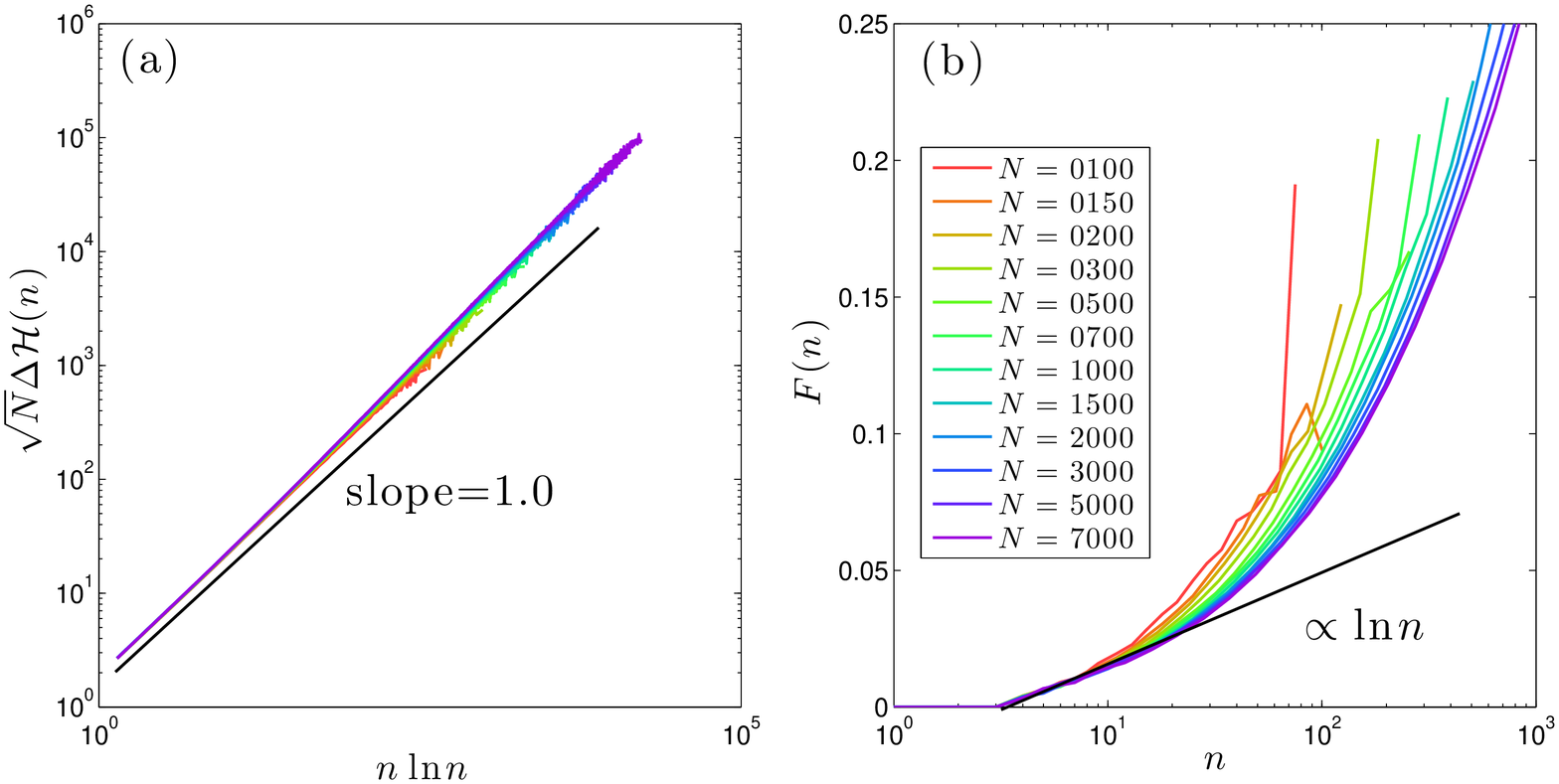}
\caption{\small{(a) The average dissipated energy $\Delta \mh$ in avalanches of size $n$ scales as $n\ln n/\sqrt{N}$. $-\Delta \mh/n$ is a measure of the typical value of the stability of most unstable spins, $\lambda_0(n)$. thus, in the thermodynamic limit, $\lambda_0\ll 1$ even for very large avalanches. (b) The average number of times, $F(n)$, spins active in avalanches of size $n$ re-flip later on in the avalanche.}}\label{propt}
\end{figure}

{\it Conclusions:}  We have studied the quasi-static dynamics in a marginally stable glass at zero temperature, focusing on a fully-connected spin glass as a model system. Our central result is that the pseudo-gap appears dynamically due to a strong frustration among the softest spins, characterized by a correlation function $C(\lambda)$ which scales inversely with the stability $\lambda$. We provided a Fokker-Planck description of the dynamics that explains  the appearance of  both the pseudo-gap and the singular correlation, and suggests a fruitful analogy between spin glass dynamics and random walks in two dimensions.  

We expect that our findings apply to other marginally stable systems, in particular hard sphere packings that display a pseudo-gap with a non-trivial exponent: $P(f)\sim f^{\theta_e}$ \cite{Wyart12,Lerner13a,Charbonneau14c,Charbonneau14} where $f$ is the contact force. Our analysis above suggests that a singular correlation function $C(f)\sim 1/f$ characterizes how all contacts are affected by the opening of a `soft' contact, i.e., by pushing two weakly-interacting particles apart (the relevant excitations in packings \cite{Wyart12,Lerner13a}). On average, contacts carrying  small forces should see their force increase by $C(f)$, a testable prediction. It also suggests a connection between the dynamics and a random walk in a space of dimension $1+\theta_e$ that will be interesting to explore further.

\begin{acknowledgements}
We thank E.~DeGiuli, J.~Lin, E.~Lerner, A.~Front for discussions. This work was supported  by the Materials Research Science and Engineering Center (MRSEC) Program of the National Science Foundation under Award number DMR-0820341, and by the National Science Foundation Grants CBET-1236378 and DMR-1105387. M.B.-J. was supported by the FPU program (Ministerio de Educaci\'on, Spain).
\end{acknowledgements}

\bibliography{Wyartbibnew}

\renewcommand{\theequation}{S\arabic{equation}}
\setcounter{equation}{0}
\setcounter{figure}{0}
\renewcommand{\thefigure}{S\arabic{figure}}

\begin{appendices}

\title{Supplementary Materials: Dynamics and Correlations among Soft Excitations in Marginally Stable Glasses}
\author{Le Yan\thanks{ly452@nyu.edu}}
\affiliation{Center for Soft Matter Research, Department of Physics, New York University, \\4 Washington Place, New York, 10003, NY}
\author{Marco Baity-Jesi\thanks{marcobaityjesi@gmail.com}}
\affiliation{Departamento de F\'isica Te\'orica I, Universidad Complutense, 28040 Madrid, Spain}
\affiliation{Dipartimento di Fisica, La Sapienza Universit\`a di Roma, 00185 Roma, Italy}
\affiliation{Instituto de Biocomputaci\'on y F\'isica de Sistemas Complejos (BIFI), 50009
Zaragoza, Spain}
\author{Markus M\"uller\thanks{markusm@ictp.it}}
\affiliation{The Abdus Salam International Center for Theoretical Physics, \\Strada Costiera 11, 34151 Trieste, Italy}
\author{Matthieu Wyart\thanks{mw135@nyu.edu}}
\affiliation{Center for Soft Matter Research, Department of Physics, New York University, \\4 Washington Place, New York, 10003, NY}

\date{\today}

\maketitle

\section{A. Stability criterion}
Consider flipping $m$ spins selected from the set of the $m'$ ($m'>m$) least stable spins. 
We make the approximation that the exchange energy of such multi-flip excitations is a random Gaussian variable when $m$ is large, as supported by Fig.~\ref{config}. The mean and the variance of $\Delta\mh$ are: 
\begin{subequations}
\begin{align}
\mu&\equiv\langle\dH\rangle=2m\left(\ml-m\mjs\right),
\label{mu}\\
\sigma^2&\equiv\langle\dH^2\rangle-\langle\dH\rangle^2=8m^2/{N}, 
\label{sigma}
\end{align}
\end{subequations}
where we have neglected the contribution from the non-diagonal terms of $\langle\sum Jss\sum Jss\rangle$. { We also omitted the fluctuations of the sum $\sum_i \lambda_i$, since that sum is anyhow always positive, and at large $m$ its fluctuations are smaller than its expectation value.} 
The maximal stability in the set of $m'$ spins is
\[
\begin{aligned}
m'&=N\int_0^{\lm}\rho(\lambda)\rd\lambda,\\
\end{aligned}
\]
and the mean values of the local stability, $\ml$, and the correlation $\mjs$ are by definition,
\[
\begin{aligned}
\ml&\equiv\int_0^{\lm}\rho(\lambda)\lambda\rd\lambda/\int_0^{\lm}\rho(\lambda)\rd\lambda,\\
\mjs&\equiv-\frac{1}{2}\int_0^{\lm}\rd\lambda\int_0^{\lm}\rd\lambda'C(\lambda,\lambda')/\lm^2.
\end{aligned}
\]
Here $C(\lambda,\lambda')$ is the correlation between the spins at the finite local stabilities $\lambda>0$ and $\lambda'>0$, defined as 
\begin{multline}
C(\lambda,\lambda')\equiv-2\langle J_{ij}s_is_j\rangle|_{\lambda_i=\lambda,\lambda_j=\lambda'}\\
\equiv \frac{1}{N^2\rho(\lambda)\rho(\lambda')}\sum_{i,j}\delta(\lambda_i-\lambda)\delta(\lambda_j-\lambda')(-2J_{ij}s_is_j).
\label{coupl}
\end{multline}
In the above $\rho(\lambda)=\sum_{i}\delta(\lambda_i-\lambda)/N$ is the  density of stabilities. 

$C(\lambda,\lambda')$ is a symmetric function, and continuous except for  the singular point $\lambda=\lambda'=0$. Defining the correlation $C(\lambda) \equiv C(\lambda,\lambda'=0)$ between a stable ($\lambda>0$) and a soft spin ($\lambda'=0$), we find that it 
 behaves as a power-law, described by Eq.~(7) in the main text. As far as scaling is concerned we thus expect $C(\lambda,\lambda')\sim C(\max[\lambda,\lambda'])$. 
 In Section B we find numerically that $C(\lambda,\lambda')\approx C(\sqrt{\lambda^2+\lambda'^2})$. 

\begin{figure}[h!]
\includegraphics[width=.8\columnwidth]{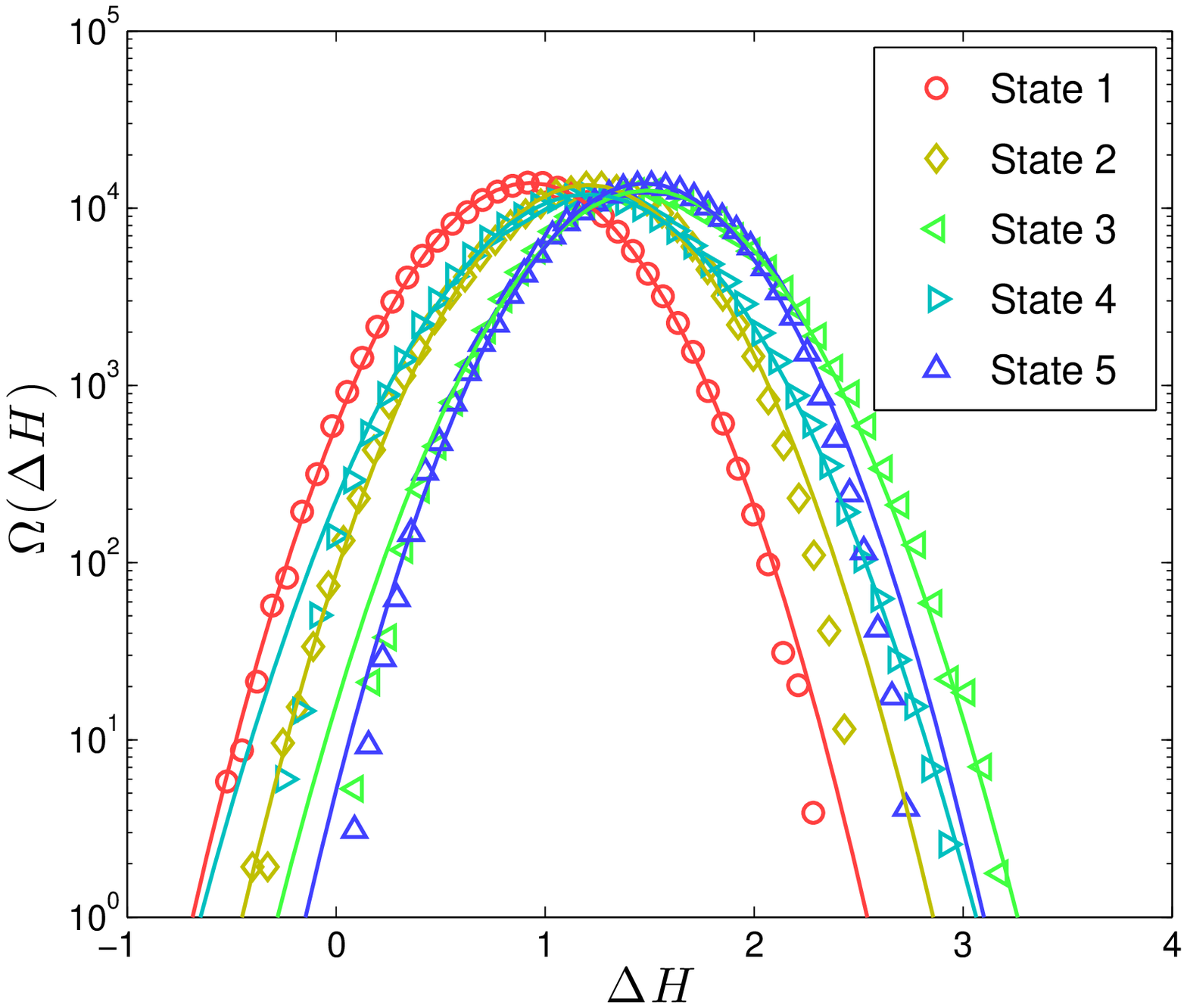}
\caption{\small{Histogram of excitations  with given energy change $\Delta H$ for different metastable states along the hysteresis curve, $m=8$, $m'=16$, and $N=3000$. }}\label{config}
\end{figure}

Assuming the pseudo-gap distribution of Eq.~(4) together with Eq.~(7) of the main text, one finds for $1\ll m<m'\ll N$:
\[
\begin{aligned}
\mu&=\frac{2m^{3/2}}{{N}^{1/(1+\theta)}}\lp a^{1/2}\frac{m'^{1/(1+\theta)}}{m^{1/2}}+\frac{b^{1/2}}{N^{\delta-(1+\gamma)/(1+\theta)}}\frac{m^{1/2}}{m'^{\gamma/(1+\theta)}}\rp,
\end{aligned}
\]
where $a$ and $b$ are numerical prefactors. { Note that the variation of $\mu$ from states to state in Fig.~\ref{config} is presumably a consequence of the small values of $m,m'$ used there.}
 
Among all excitations of $m$ out of $m'$ spins, the number of sets that lower the total energy, $\Delta\mh<0$, is
\[
\Omega(m,m')=\binom{m'}{m}\Phi\lp\frac{\mu}{\sqrt{2}\sigma}\rp\approx\frac{m'!}{m!(m'-m)!}\exp\lp-\frac{\mu^2}{2\sigma^2}\rp,
\]
where $\Phi$ is the complementary error function.

We define a free energy as the logarithm of $\Omega$, 
\begin{eqnarray}
\label{free}
&&f(r)\equiv-\frac{1}{m}\ln\Omega\sim -r\ln r+(r-1)\ln(r-1) \\
&&+\frac{a}{4}r^{2/(1+\theta)}\lp\frac{m}{N}\rp^{\frac{1-\theta}{1+\theta}}+\frac{b}{4}r^{-2\gamma/(1+\theta)}\frac{m^{1-2\gamma/(1+\theta)}}{N^{2\delta-1-2\gamma/(1+\theta)}},\nonumber
\end{eqnarray}
where $r\equiv m'/m$. 
{The cross term ($\sim \langle\lambda\rangle_{m'} \langle Jss\rangle_{m'})$ in $\mu^2$ has been neglected, as it cannot diverge faster than the terms in the second line of Eq. (\ref{free}).} A positive $f(r)$ implies no unstable excitations of the initial state in the limit $m\gg 1$. Stability is thus achieved if either term in the second line of Eq.~(\ref{free}) diverges as $m\to \infty$. This requires that either $\theta\geq1$, or $\gamma\leq1$ and $\delta\leq1$, as explained in the main text. 

\begin{figure}[h!]
\includegraphics[width=0.8\columnwidth]{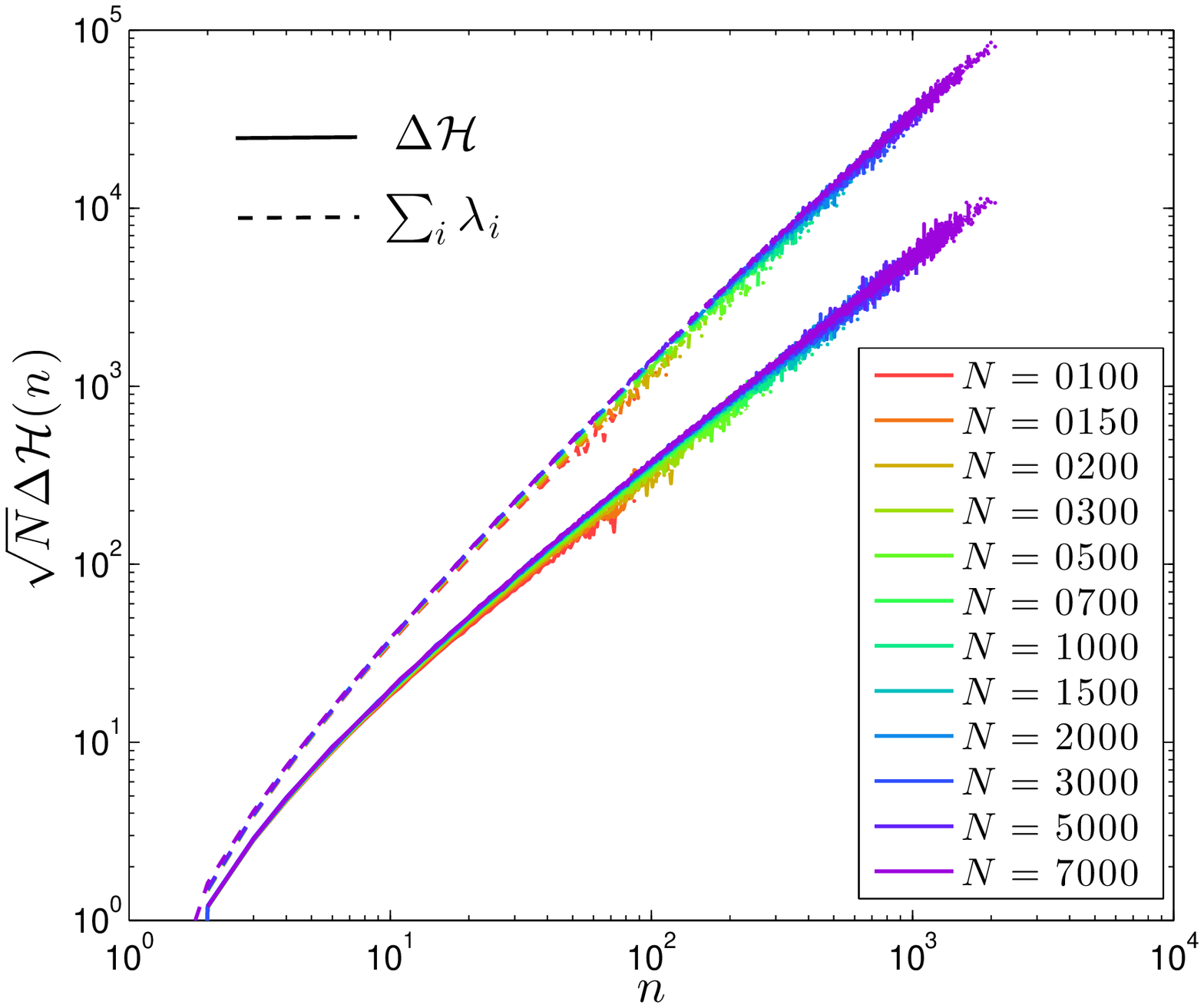}
\caption{ \small{ { The lower data set in solid lines is} the total energy  $\Delta \mh(n)$  dissipated in avalanches of size $n$. { The upper data set in dashed lines}  is the sum of local stabilities (before the avalanche) of spins that are going to flip in the avalanche, $\sum_{i\,  {\rm flip}}\lambda_i$. This shows that the dissipated energy is vanishingly small as compared to the na\"ive sum over local  stabilities, as $n\to\infty$, since the two curves scale as different power laws with $n$. 
}}\label{dee}
\end{figure}

In the marginal case, $\theta=\gamma=\delta=1$, 
the free energy becomes: 
\be
\label{freeb}
f(r)\approx -r\ln r+(r-1)\ln(r-1)+\frac{ar^2+b+2\sqrt{ab}r}{4r}.
\ee
An interesting finding is that $f(r)$ is minimized by a finite ratio $r^*$, independently of the set size $m$. $r^*$  is an estimate of the optimal volume $m'^*= r^* m$ for finding energy lowering subsets of size $m$. 

The fact that stability is controlled by the ratio $r$  instead of the absolute values of $m$ or $m'$ is consistent with the observation that the dynamics proceeds via power-law avalanches with no scales (a fact implied by the argument of Ref.~\cite{Muller14}). Indeed in the marginal case multi-flip excitations can be slightly unstable (as illustrated in Fig.~\ref{config}), and can be triggered as the magnetic field is increased. Marginality is apparent when analyzing these avalanches: We find that the energy dissipated in avalanches is much smaller  than the na\"ive estimate which sums all local stabilities of spins that are going to flip in the avalanche, $\sum_{i\,  {\rm flip}}\lambda_i$. This reflects the fact that the total energy change in the avalanche, $\Delta\mh$, is a result of a near cancellation of several terms, as discussed in the main text, verified in Fig.~\ref{dee}. 

%

\section{B. Two-point correlation}

 We numerically measured the two-point correlation, and found $NC(\lambda,\lambda')=D/\sqrt{[\lambda+c(N)]^2+[\lambda'+c(N)]^2}$ as long as $\lambda,\lambda'\ll1$. Here, $c(N)=1.1\sqrt{\ln N/N}$ is a finite size correction which vanishes in the thermodynamic limit. This result is illustrated in Fig.~\ref{correl}. When finite size effects are negligible this implies that $C(\lambda,\lambda')\approx  C(\sqrt{\lambda^2+\lambda'^2)}$. 

\begin{figure}[h!]
\includegraphics[width=.48\columnwidth]{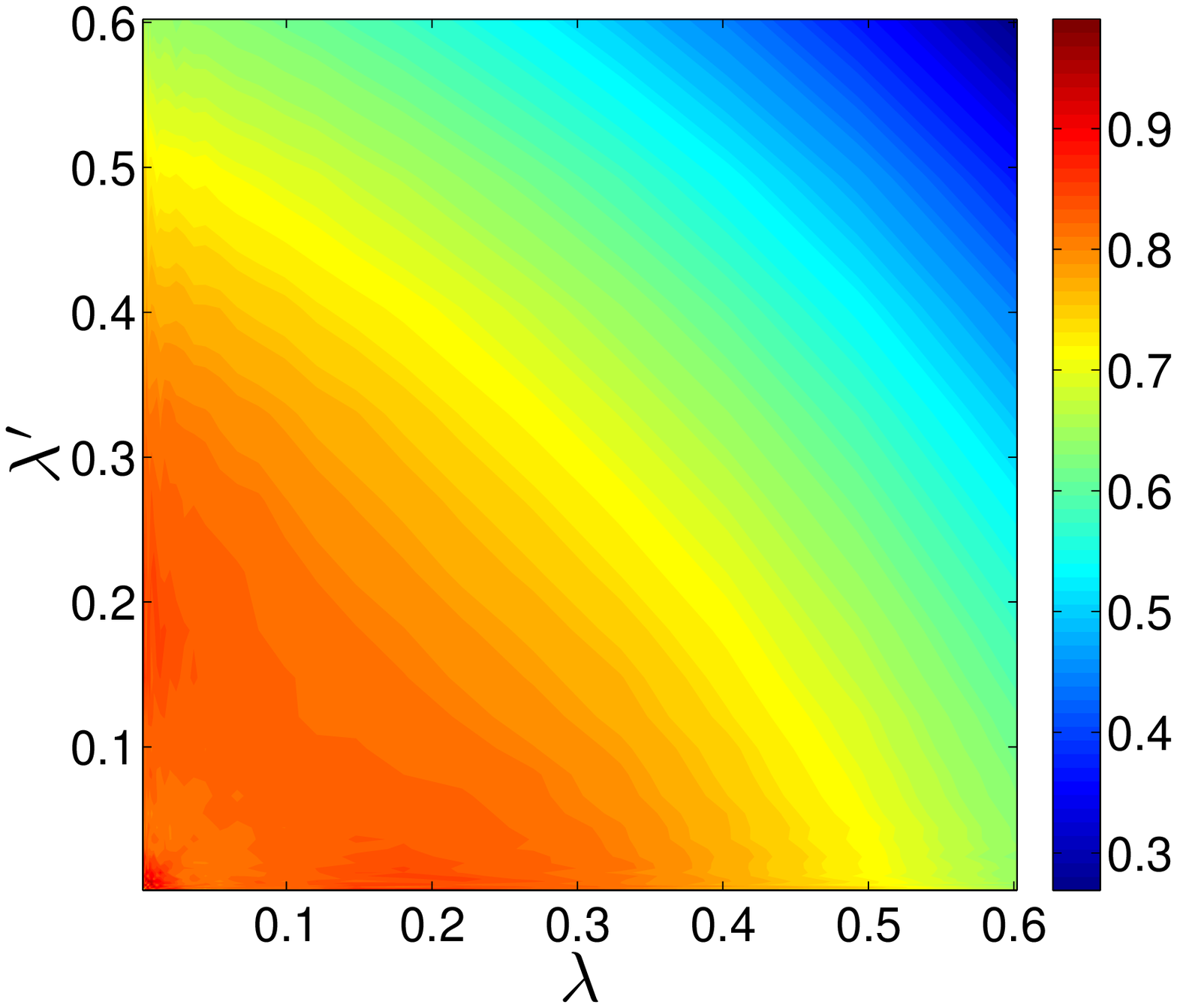}
\includegraphics[width=.48\columnwidth]{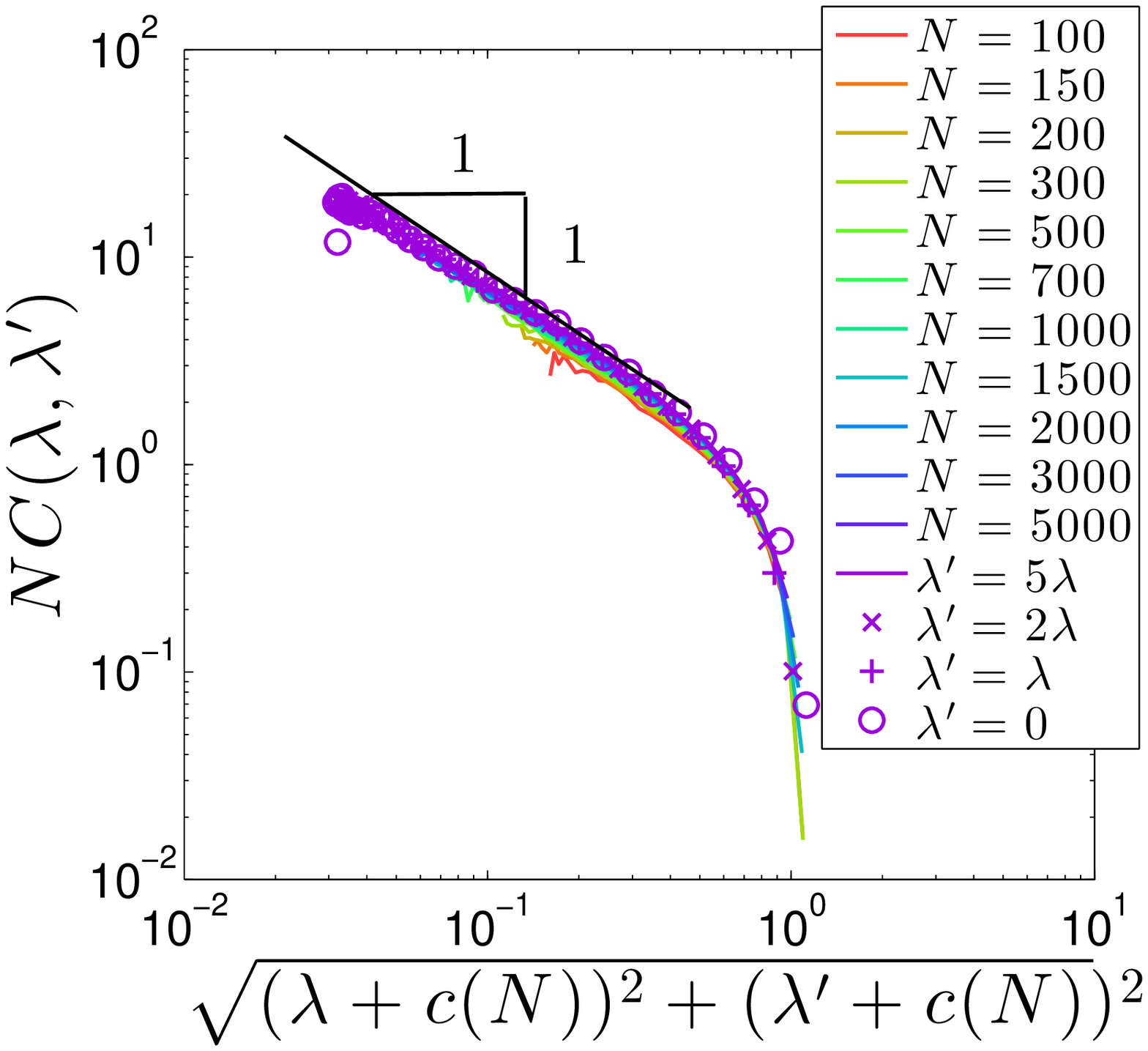}
\caption{\small{Left: the quantity $NC(\lambda,\lambda')\sqrt{(\lambda+c(N))^2+(\lambda'+c(N))^2}$ is numerically computed for various $\lambda$ and $\lambda'$, and behaves nearly as a constant (as the color code indicates, this quantity only changes by a factor 3 in the entire range considered. Right: Correlation $C(\lambda,\lambda')$, for $\lambda'=5\lambda$ with different system sizes $N$ and for different directions $\lambda'=a\lambda$, $a=0,1,2,5$ with $N=5000$.}}\label{correl}
\end{figure}


\section{C. Dynamical Constraints on $\theta$}

In the main text we  argue that the density of local stabilities $\rho(\lambda)$ satisfies a Fokker-Planck (FP) equation of the type:
\be
\label{001}
\partial_t\rho(\lambda,t)=-\partial_{\lambda}\,\left[v(\lambda,t)-\partial_{\lambda}D(\lambda,t)\right]\rho(\lambda,t),
\ee
with a reflecting boundary at $\lambda=0$. In addition, we show that  correlations emerge dynamically, which in the steady state take the form:
\be
\label{002}
C(\lambda)N=v_{\rm ss}(\lambda)=D\partial_{\lambda}\rho_{\rm ss}(\lambda)/\rho_{\rm ss}(\lambda).
\ee
{These results seem to imply that the fact that correlations are necessary to obtain a steady state, is not constraining the latter in any way. Indeed, any function $\rho_{\rm ss}(\lambda)$ could in principle appear as a steady state, as long as the correlations satisfy Eq.~(\ref{002}).} This is true in particular for any scaling function $\rho_{\rm ss}(\lambda)\sim \lambda^\theta$, whatever the value of $\theta$. However, we now argue that only the case $\theta=1$ is a viable solution in SK model. 

{\it Excluding $\theta<1$:}  Our FP description only applies beyond the discretization scale of the kicks due to flipping spins, which are of order $J\sim 1/\sqrt{N}$. 
In particular, from its definition, $C(\lambda)$ must be bounded by $1/\sqrt{N}$. Taking this into account, Eq.~(\ref{002}) should be modified to:
\be
\label{003}
v_{\rm ss}(\lambda)\approx \min\{D\partial_{\lambda}\rho_{\rm ss}(\lambda)/\rho_{\rm ss}(\lambda)\sim 1/\lambda, \sqrt{N}\}.
\ee
This modification has no effect when $\theta\geq 1$, since in that case $\lambda_{\min}\sim N^{-1/(1+\theta)}\geq 1/\sqrt{N}$. 
In contrast, pseudo-gaps with $\theta<1$ have $\lambda_{\min}\ll 1/\sqrt{N}$. To maintain such a pseudo-gap in a stationary state, one would require correlations much larger than what the discreteness of the model allows. Pseudo-gaps with $\theta<1$ are thus not admissible solutions of Eqs.~(\ref{001},~\ref{003}).

%

{\it Excluding $\theta>1$:}
In this case, $\lambda_{\min}\gg 1/\sqrt{N}\sim J$. Thus when one spin flips, the second least stable spin will not flip in general, and avalanches are typically of size unity \cite{Muller14}. It can easily be shown that in that case, our assumption (ii) in the main text is violated:  
the number of flips per spin along the loop would be small (in fact it would even vanish in the thermodynamic limit, which is clearly impossible). In terms of our FP description, the motion of the spin stabilities due to other flips would be small in comparison with the motion of the stabilities inbetween avalanches, due to changes of the magnetic field.   Making the crude assumption that the magnetization is random for any $\lambda$, the change of external magnetic field leads to an additional diffusion term in the Fokker-Planck equation:
\be
\label{fpwithd}
\partial_t\rho(\lambda,t)=-\partial_{\lambda}(v-D\partial_{\lambda})\rho(\lambda,t)+D_h\partial_{\lambda}^2\rho(\lambda,t),
\ee
where the term $D_h$ is related to the typical field increment $h_{\rm min}\sim \lambda_{\rm min}$ required  to trigger an avalanche. Indeed $D_{h}\sim Nh_{\rm min}^2\sim N^{(\theta-1)/(\theta+1)}\gg D\sim 1$. Under these circumstances, Eq.~(\ref{002}) does not hold. The dynamics would be a simple diffusion with reflecting boundary, whose only stationary solution corresponds to $\theta=0$, violating our hypothesis $\theta>1$.  

\section{D. Analogy to a $d$ dimensional random walk}
Consider a non-biased random walk in $d$ dimension,
\be
\label{drw}
\vec{x}(t+\rd t) = \vec{x}(t) + \sqrt{{2D}\rd t}\vec{\eta}(t),
\ee
where $D$ is the diffusion constant, and $\vec{\eta}$ is a random Gaussian vector. Then the probability density $P(\vec{x},t)$ satisfies the Fokker-Planck equation~\cite{Risken96},
\[
\partial_{t}P(\vec{x},t) = D\nabla^2P(\vec{x},t).
\]

The process is angle-independent, so $P(\vec{x},t)$ satisfies, 
\be
\label{bfp}
\partial_tP(r,t) = D\partial_r^2P(r,t)+D\frac{d-1}{r}\partial_rP(r,t),
\ee
where $r=|\vec{x}|$.
However, $P(r,t)$ is not a probability density with respect to $r$. Including the Jacobian of the change of variables, the probability density, $\rho(r,t)\equiv\Omega_dr^{d-1}P(r,t)$, satisfies the corresponding Fokker-Planck equation, 
\be
\label{ffp}
\partial_t\rho(r,t) = -\partial_r\lp D\frac{d-1}{r}-D\partial_r\rp\rho(r,t),
\ee
with $\Omega_d$ the solid angle in $d$ dimensions. This Fokker-Planck equation 
for the radial component of a $d$-dimensional unbiased random walk is exactly the same as the diffusion equation, Eq.~(11), with reflecting boundary at $\lambda=0$ and diverging drift $v(\lambda)=\theta/\lambda$ at small $\lambda$. The case of pseudo-gap exponent $\theta=1$ in the spin model thus maps to a two-dimensional random walk.

\end{appendices}

\end{document}